\documentclass[english,prl,aps,twocolumn,fleqn,epsfig,showpacs]{revtex4}
\usepackage[T1]{fontenc}
\usepackage[latin9]{inputenc}
\usepackage{babel}

\usepackage{amsmath}
\usepackage{graphicx}
\usepackage{amssymb}
\usepackage[unicode=true, pdfusetitle,
 bookmarks=false,
 breaklinks=false,pdfborder={0 0 1},backref=false,colorlinks=false]
 {hyperref}

\makeatletter

\providecommand{\LyX}{L\kern-.1667em\lower.25em\hbox{Y}\kern-.125emX\@}

\@ifundefined{textcolor}{}
{%
 \definecolor{BLACK}{gray}{0}
 \definecolor{WHITE}{gray}{1}
 \definecolor{RED}{rgb}{1,0,0}
 \definecolor{GREEN}{rgb}{0,1,0}
 \definecolor{BLUE}{rgb}{0,0,1}
 \definecolor{CYAN}{cmyk}{1,0,0,0}
 \definecolor{MAGENTA}{cmyk}{0,1,0,0}
 \definecolor{YELLOW}{cmyk}{0,0,1,0}
 }

\makeatother

\begin{document}

\title{All Possible Coupling Schemes in $XY$ Spin Chains for Perfect State
Transfer}

\author{Yaoxiong Wang}

\affiliation{Institute of Intelligent Machines, Chinese Academy of Sciences, Hefei,
China 230031}

\author{Herschel Rabitz}

\affiliation{Department of Chemistry, Princeton University, Princeton, NJ 08544}

\author{Feng Shuang}

\affiliation{Institute of Intelligent Machines, Chinese Academy of Sciences, Hefei,
China 230031}
\begin{abstract}
We investigate quantum state transfer in $XY$ spin chains and propose
a recursive procedure to construct the nonuniform couplings of these
chains with arbitrary length to achieve perfect state transfer(PST).
We show that this method is capable of finding all possible coupling
schemes for PST. These schemes, without external control fields, only
involve preengineered couplings but not dynamical control of them,
so they can be simply realized experimentally. The analytical solutions
provide all information for coupling design.
\end{abstract}

\pacs{03.67.Hk, 05.50.+q}

\maketitle

Quantum information and quantum computation can process lots of tasks
which are intractable with classical technologies.
Although many schemes such as quantum dots\cite{Loss98PRA}, ion trap\cite{Cirac95PRL},
NMR\cite{Cory97PANS} have been discussed extensively, a macroscopic
scalable quantum computer still seems to need a channel, often known
as quantum wire, to transmit or exchange quantum states between inner
parts of the quantum computer. These architectures require to implement
a transmission process for an unknown quantum state from one place
to another which is often called quantum state transfer. In a seminal
paper\cite{Bose03PRL}, Bose proposed a spin chain model, whose evolution
was governed by a reasonable Hamiltonian, and considered the fidelity
of state transfer in this model. Similar results were also derived
by studying dynamical properties of entanglement transition in Heisenberg
$XY$ spin chain\cite{Subrahmanyam04PRA}. This model, in which two
processors are connected through a spin chain as quantum wire, is
useful for quantum computation based on Heisenberg interaction\cite{DiVincenzo00Nature}
or measurements\cite{oneway}. 

Although some important and significant results have been found, see
for example\cite{Bose03PRL,Subrahmanyam04PRA}, all of the available
results are just concerned with uniform interaction, i.e. the couplings
between any two nearest-neighbor sites are the same. For this case,
however, it is shown that when $N\geq4$, where $N$ is the number
of the sites in $XY$ chain, PST is impossible\cite{Christandl}.
This drawback of uniform interaction motivates people to find some
modified models to achieve "long" distance PST. Some works considered
long-range interactions\cite{longint}, and some concentrated
on numerical simulations\cite{simu}. One feasible
choice is to preengineer the couplings\cite{Christandl}, i.e.
choose special nonuniform couplings to achieve PST, and some specific
analytical coupling schemes were found\cite{Christandl,Albanese04PRL,Sun05PRA}.
The necessary and sufficient conditions for the couplings of PST,
which can provide a criterion to verify the preengineered schemes
as well as to find new analytical ones, were derived from a more systematical
treatment of this problem\cite{Bose05PRA} by mirror inversion\cite{Albanese04PRL}
and quantum computation. However, all these preengineered schemes
are obtained through a "verifiable" but not "constructive"
way. Thus, we have not yet got all possible coupling schemes for PST.

In this Letter, we start from the necessary and sufficient conditions
of PST. After preselecting the eigenvalues of a $XY$ spin chain Hamiltonian,
we propose two recursive formulas of the couplings for both even and
odd $N$ cases and prove them by mathematical induction. Further discussions
demonstrate that this method is capable of finding all possible coupling
schemes for PST in $XY$ chain with arbitrary length. Experimentally,
our PST schemes can be realized, for example, by superconducting circuits
and quantum bus\cite{SCandQB}, nanoelectromechanical
resonator arrays\cite{Rabl10Nature} or cold-atom optical lattice\cite{Bloch08Nature}.

Next, we first review some basic concepts of state transfer protocol
using spin chain as the channel\cite{Bose03PRL,review}.
An unknown qubit, as encoded in site $1$, is attached to one end
of a spin chain when the chain is initialized to the all spin-down
ground state (state initialization is not necessary\cite{DiFranco08PRL},
and our results can be generalized to these cases). Due to the coupling
between site $1$ and $2$, free evolution of the system causes the
unknown state to distribute among the chain. After a specific interval,
we want to recover this unknown state at the opposite end of the chain
to achieve state transfer. 

A reasonable Hamiltonian for this task is $XY$ type Hamiltonian\begin{equation}
H=\frac{1}{2}\sum_{i=1}^{N-1}J_{i}(\sigma_{i}^{x}\sigma_{i+1}^{x}+\sigma_{i}^{y}\sigma_{i+1}^{y})-\frac{1}{2}\sum_{i=1}^{N}B_{i}(\sigma_{i}^{z}-1),\label{XY Hamiltonian}\end{equation}
where $J_{i}$ is the coupling strength between sites $i$ and $i+1$,
and $B_{i}$ is the external static potential, or control field, at
site $i$. $\sigma^{x},\sigma^{y},\sigma^{z}$ are the three Pauli
matrices. One important observation is that Hamiltonian\eqref{XY Hamiltonian}
commutates with the total z-spin operator $\sum_{i=1}^{N}\sigma_{i}^{z}$.
Thus, $\sum_{i=1}^{N}\sigma_{i}^{z}$ is a conservation, and the evolution
of the system in these state transfer cases will just involve the
subspace spanned by ground state and $N$ one-site excited states.
By the Jordan-Wigner transform, which maps\eqref{XY Hamiltonian}
to

\begin{equation}
H=\sum_{i=1}^{N-1}J_{i}(a_{i}^{\dagger}a_{i+1}+a_{i+1}^{\dagger}a_{i})+\sum_{i=1}^{N}B_{i}a_{i}^{\dagger}a_{i},\label{Hamiltonian}\end{equation}
$XY$ model can be solved exactly. Hamiltonian\eqref{Hamiltonian}
describes an $N$-site hopping model subjects to nonuniform external
fields. Let $\left|i\right\rangle $ denotes the single excited state
at site $i$, Hamiltonian\eqref{Hamiltonian} in a $2^{N}$-dimensional
space will reduce to an $N$-dimensional subspace spanned by $\{\left|i\right\rangle \}$.
Explicitly,\begin{equation}
H_{N}=\left(\begin{array}{ccccc}
B_{1} & J_{1} & 0 & \cdots & 0\\
J_{1} & B_{2} & J_{2} & \cdots & 0\\
0 & J_{2} & B_{3} & \cdots & 0\\
\vdots & \vdots & \vdots & \ddots & J_{N-1}\\
0 & 0 & 0 & J_{N-1} & B_{N}\end{array}\right)\label{HMatrix}\end{equation}
in $\{\left|i\right\rangle \}$ basis. The fidelity of this transfer
procedure can be expressed as $\left\langle N\right|e^{-iH_{N}\tau}\left|1\right\rangle $,
where $\tau$ is the time interval of the free evolution. The equivalent
conditions for PST, i.e. $\left|\left\langle N\right|e^{-iH_{N}\tau}\left|1\right\rangle \right|=1$,
are: (a) the reflection symmetry $B_{i}=B_{N+1-i}$ and $J_{i}=J_{N-i}$.
(b) after sorting the eigenvalues of $H_{N}{\displaystyle \frac{\tau}{\pi}}$
in decreasing order, the difference between any two adjacent eigenvalues
is an odd number\cite{Bose05PRA}. All schemes discovered before required
(a) as part of their protocols and designed the eigenvalues of $H_{N}{\displaystyle \frac{\tau}{\pi}}$
to be $\{-k,-k+1,\cdots k-1,k\}$ for $2k\in\mathbb{N}$\cite{Christandl,Albanese04PRL},
$\{q(k^{2}+k)+(2p+1)k\}$ for $k=0,\ldots,N$\cite{Albanese04PRL}
or $\{-k+\frac{1}{2}-n,\ldots,-k-\frac{3}{2},-k-\frac{1}{2},k+\frac{1}{2},k+\frac{3}{2},\ldots,k-\frac{1}{2}+n\}$\cite{Sun05PRA}
which all satisfy (b). All these coupling schemes are special solutions
for PST, and our main result in this Letter is to show how to get
all possible couplings for PST in the absence of external fields,
i.e. $B_{i}=0$. Because of the perfect transfer condition (a) and
the postulation $B_{i}=0$, Hamiltonian\eqref{HMatrix} becomes \begin{equation}
H_{N}=\left(\begin{array}{cccccc}
0 & J_{1} & 0 & \cdots & 0 & 0\\
J_{1} & 0 & J_{2} & \cdots & 0 & 0\\
0 & J_{2} & 0 & \cdots & 0 & 0\\
\vdots & \vdots & \vdots & \ddots & J_{2} & 0\\
0 & 0 & 0 & J_{2} & 0 & J_{1}\\
0 & 0 & 0 & 0 & J_{1} & 0\end{array}\right)\label{HN}\end{equation}
whose eigenvalues are symmetric about zero. Owing to this symmetry,
there are only $[N/2]$ independent couplings and $[N/2]$ independent
eigenvalues in\eqref{HN} ($0$ is always an eigenvalue when $N$
is odd). Our purpose is to construct the couplings $\{J_{i}\}$ from
a set of preselected eigenvalues $\{\Lambda_{i}\}$ satisfy (b). We
will first consider even $N$ cases and show how to derive $\{J_{i}\}$
effectively. Then, we generalize these results to odd $N$ cases,
and finally show the completeness of this method, i.e. it can get
all possible coupling schemes for PST. 

For even $N$ cases, we assume the eigenvalues of $H_{N}$ are $\{\pm\Lambda_{1},\ldots,\pm\Lambda_{n}\}$
where $n=N/2$ , $\Lambda_{i}\in\mathbb{N}$ and $\Lambda_{1}>\Lambda_{2}>\ldots>\Lambda_{n}>0$
(if none of $\{J_{i}\}$ is zero, then the eigenvalues of $H_{N}$
are nondegenerate\cite{Wilkinson}), and omit the scale factor ${\displaystyle \frac{\tau}{\pi}}$.
$\{J_{i}\}$ and $\{\Lambda_{i}\}$ are connected through the characteristic
polynomial of the Hamiltonian\eqref{HN}: \begin{equation}
\text{Det}(H_{N}-\lambda I)=\prod_{i=1}^{n}(\lambda^{2}-\Lambda_{i}^{2})\label{CharPoly}\end{equation}
which, by expanding it with respect to $\lambda^{2}$, is equivalent
to a set of equations:\begin {subequations}\label{6}\begin{eqnarray}
\sum_{i=1}^{N-1}J_{i}^{2} & = & \sum_{i=1}^{n}\Lambda_{i}^{2}\label{6.a}\\
 & \vdots\nonumber \\
\sum_{k_{i+1}-k_{i}\geq2}J_{k_{1}}^{2}\cdots J_{k_{n}}^{2} & = & \sum_{k_{i+1}>k_{i}}\Lambda_{k_{1}}^{2}\cdots\Lambda_{k_{n}}^{2},\\
 & \vdots\nonumber \\
\prod J_{1}^{2}J_{3}^{2}\ldots J_{N-1}^{2} & = & \prod_{i=1}^{n}\Lambda_{i}^{2}\end{eqnarray}
\end {subequations}and we want to derive $\{J_{i}\}$ from $\{\Lambda_{i}\}$.
This is often called an inverse problem. Notice that we still use
$J_{N-i}$ rather than $J_{i}$ when $i\leq N/2$ despite they are
equal just for elegance of the expressions. We first introduce some
notations for convenience. Denote $j_{n}^{[N]}=J_{n}^{[N]}$ and $j_{i}^{[N]}=(J_{i}^{[N]})^{2}$
for $1\leq i\leq n-1$ whose meaning will become clear soon. Here,
the superscript $N$ denotes the dimension of the matrix $H_{N}$
and we imply the eigenvalues of $H_{D}$ are $\{\pm\Lambda_{1},\dots,\pm\Lambda_{D/2}\}$
and its couplings are $\{J_{i}^{[D]}\}$. The main idea is to obtain
$\{j_{i}^{[D]}\}$ from $\{j_{i}^{[D-2]}\}$ when we require the Hamiltonians
construct by them respectively share the same eigenvalues $\{\pm\Lambda_{1},\dots,\pm\Lambda_{(D-2)/2}\}$.
Further, denote $\Gamma_{k}^{N}=\frac{j_{n-1}^{[N-2]}j_{n-2}^{[N-2]}\cdots j_{k-1}^{[N-2]}}{j_{n}^{[N]}j_{n-1}^{[N]}\cdots j_{k+1}^{[N]}}$
for $1\leq k<n$, where $j_{0}^{[D]}\equiv0$, and $\Gamma_{n}^{N}=j_{n-1}^{[N-2]}$.
Denote $\Delta_{k}^{N}=\frac{j_{k}^{[N-2]}j_{k+2}^{[N-2]}j_{k+4}^{[N-2]}\cdots}{j_{k+2}^{[N]}j_{k+4}^{[N]}j_{k+6}^{[N]}\cdots}$
and $\Delta_{n-1}^{N}=j_{n-1}^{[N-2]}$ , $\Delta_{n}^{N}=1$ where
the products in the numerators and denominators involve terms only
if the indices of them are not larger than $n-1$ and $n$ respectively.
With these notations, we will show the following equation permits us
to get $\{J_{i}\}$ from $\{\Lambda_{i}\}$ directly:

\begin{eqnarray}
j_{i}^{[N]} & = & \Gamma_{i}^{N}-(-1)^{i}\Lambda_{n}\Delta_{i}^{N}\;\;\; i=1,\cdots,n.\label{EvenRec}\end{eqnarray}
Eq.\eqref{EvenRec} allows to construct $j_{i}^{[N]}$ from $\{j_{n-1}^{[N-2]},\ldots,j_{i-1}^{[N-2]}\}$
,$\{j_{n}^{[N]},\cdots,j_{i+1}^{[N]}\}$ and $\Lambda_{n}$. Thus,
when we know $\{j_{i}^{[N-2]}\}$, by adding one more parameter $\Lambda_{n}$,
we can derive $j_{n}^{[N]},j_{n-1}^{[N]},\cdots j_{1}^{[N]}$ one
by one explicitly. Now, we need to prove Eq.\eqref{EvenRec} is consistent
with Eq.(\ref {6}). Direct calculation shows $\Lambda_{n}$ satisfies
a continued fraction:\begin{equation}
\frac{j_{n}^{[N]}}{\frac{j_{n-1}^{[N]}}{\frac{\vdots}{\frac{j_{1}^{[N]}}{\Lambda_{n}}-\Lambda_{n}}+\Lambda_{n}}+(-1)^{n-1}\Lambda_{n}}=1.\label{ConFra}\end{equation}
Eq.\eqref{ConFra} is equivalent to $\text{Det}(H_{N}-\Lambda_{n}I)=0$.
Actually, by expanding Det$(H_{i}-\Lambda I)$ in terms of order $i-1$
determinants, we will find the original continued fraction for $\text{Det}(H_{N}-\Lambda_{n}I)=0$
is\begin{equation}
\frac{(J_{N}^{[N]})^{2}}{\frac{(J_{N-1}^{[N]})^{2}}{\frac{\vdots}{\frac{(J_{2}^{[N]})^{2}}{\frac{(J_{1}^{[N]})^{2}}{\Lambda_{n}}-\Lambda_{n}}+\Lambda_{n}}-\Lambda_{n}}+(-1)^{N-1}\Lambda_{n}}+(-1)^{N}\Lambda_{n}=0.\label{ConFraOri}\end{equation}
Due to the symmetry between $J_{i}$ and $J_{N-i}$, we can move upper
half of the continued fraction to the right hand of the equal sign.
After taking a square root on both sides, we obtain Eq.\eqref{ConFra},
which means $\Lambda_{n}$ is actually an eigenvalue of \eqref{HN}.
The square root operation is exactly the origin of why we denote $j_{n}^{[N]}=J_{n}^{[N]}$
but $j_{i}^{[N]}=(J_{i}^{[N]})^{2}$ for $i\neq n$ before. Next,
we will prove Eq.\eqref{EvenRec} is correct for arbitrary $N$ by
mathematical induction. We assume the permutations of $\{\Lambda_{1},-\Lambda_{2},\Lambda_{3},\ldots,(-1)^{n}\Lambda_{n-1}\}$
form a group keeps $\{j_{i}^{[N-2]}\}$ unchanged which is actually
true for $\{j_{i}^{[4]}\}$. The following step is to prove $\{j_{i}^{[N]}\}$
are also invariant under the permutation of $\Lambda_{n}$ and $-\Lambda_{n-1}$which,
with the assumption above, directly induces $\{\pm\Lambda_{i}\}$
for $i=1,\ldots,n$ are the eigenvalues of $H_{N}$ when $\{J_{i}^{[N]}\}$
are constructed from Eq.\eqref{EvenRec}. Obviously, $j_{n}^{[N]}$
is unchanged under the permutation of $\Lambda_{n}$ and $-\Lambda_{n-1}$.
For $j_{n-1}^{[N]}$, we can expand it using $\Lambda_{n}$, $\Lambda_{n-1}$
and $\{j_{i}^{[N-4]}\}$ in which $\{j_{i}^{[N-4]}\}$ are irrelevant
to $\Lambda_{n}$ and $\Lambda_{n-1}$. Notice that this expression
has similar form with $j_{n-5}^{[N-2]}$ when $j_{n-5}^{[N-2]}$ is
expanded by $\Lambda_{n-1}$, $\Lambda_{n-2}$ and $\{j_{i}^{[N-6]}\}$,
and if we replace $\Lambda_{n}$, $\Lambda_{n-1}$ and $\{j_{i}^{[N-4]}\}$
in $j_{n-1}^{[N]}$ by $-\Lambda_{n-1}$, $-\Lambda_{n-2}$ and $\{j_{i}^{[N-6]}\}$
respectively, we will find they are indeed the same one. Owing to
the assumption that the permutation of $\Lambda_{n-1}$ and $-\Lambda_{n-2}$
keeps $j_{n-5}^{[N-2]}$ unchanged, we conclude that $j_{n-1}^{[N]}$ is
also unchanged under the permutation of $\Lambda_{n}$ and $-\Lambda_{n-1}$.
This method, demonstrating the invariance by replacement, is applicable
for other $j_{i}^{[N]}$, and we can further prove all $\{j_{i}^{[N]}\}$
are invariant under the permutation of $\Lambda_{n}$ and $-\Lambda_{n-1}$. 

\begin{figure}[b]
\includegraphics[clip,scale=0.7]{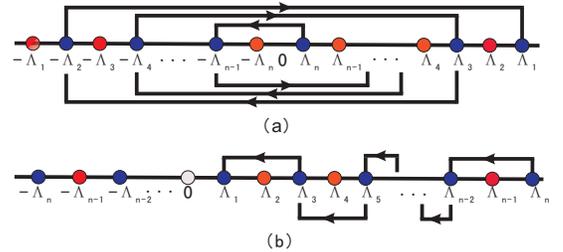}\caption{(a) and (b) are recursive procedures for even and odd $N$ cases respectively.
Permutations of $\{\Lambda_{i}\}$ with the same colour form one group.
Although $0$ is always an eigenvalue in (b), the permutation group
does not contain it. Arrows on the straight lines indicate the recursive
directions, e.g. the rightest arrow in (b) implies if $\Lambda_{n}$
is an eigenvalue of \eqref{HN} when $\{J_{i}\}$ are constructed
from Eq.(\ref {10}) then so does $\Lambda_{n-2}$.}\label{fig1}
\end{figure}

Combining this proof and the fact that $\{\Lambda_{1},-\Lambda_{2}\}$
actually form a group for $\{j_{1}^{[4]},j_{2}^{[4]}\}$, we can prove
the permutations of $\{\Lambda_{1},-\Lambda_{2},\ldots,(-1)^{n+1}\Lambda_{n}\}$
form a group keeps $\{j_{i}^{[N]}\}$ unchanged. Furthermore, if $\Lambda_{n}$
is an eigenvalue of \eqref{HN}, then, according to Eq.\eqref{ConFra}, $\{\pm\Lambda_{i}\}$ for $i=1,\ldots,n$
are all eigenvalues of \eqref{HN}. Fig.\eqref{fig1} shows idea of the proof.

For odd $N$ cases, we assume $\Lambda_{n}>\Lambda_{n-1}>\ldots>\Lambda_{1}>0$.
Define $j_{n}^{[N]}=2(J_{n}^{[N]})^{2}$ and $j_{i}^{[N]}=(J_{i}^{[N]})^{2}$
for $1\leq i\leq n-1$, $n=(N-1)/2$, and define $\Gamma_{k}^{N}$
and $\Delta_{k}^{N}$ as the same as even $N$ cases. The corresponding
recursive formula for odd $N$ is\begin {subequations}\label{10}\begin{eqnarray}
j_{k}^{[N]} & = & \Lambda_{n}^{2}\Delta_{k}^{N}-\Gamma_{k}^{N}\label{10.a}\\
j_{k-1}^{[N]} & = & \Delta_{k-1}^{N}-\Gamma_{k-1}^{N}\end{eqnarray}
\end {subequations}where $k=n,n-2,n-4,\cdots$ till we get $j_{1}^{[N]}$. The difference between
$n$ and $k$ in Eq.\eqref{10.a} is an even number which implies
$\Lambda_{n}^{2}$ appears in the right hand of Eq.(\ref {10}) alternately.
Just like even $N$ cases, we can directly check $\Lambda_{n}$ is
an eigenvalue of \eqref{HN} by the continued fraction representation
when $\{J_{i}\}$ are expressed by Eq.(\ref {10}), and the factor
$2$ appears in the definition of $j_{n}^{[N]}$ also comes from the
continued fraction structure. Although the main idea is the same,
there are still some differences between even and odd cases. First,
$\{j_{i}^{[N]}\}$ are no longer unchanged under the permutation of
$\Lambda_{n}$ and $-\Lambda_{n-1}$ when $N$ is odd. Instead, the
permutations of $\{\Lambda_{n},\Lambda_{n-2},\Lambda_{n-4},\ldots\}$
and $\{\Lambda_{n-1},\Lambda_{n-3},\Lambda_{n-5},\ldots\}$ form two
groups keep $\{j_{i}^{[N]}\}$ invariant respectively (if we consider
all the eigenvalues $\{\pm\Lambda_{i}\}$, then both even and odd
cases have two groups formed by interlaced eigenvalues respectively
which keep $\{j_{i}^{[N]}\}$ unchanged, see Fig.\eqref{fig1}). Second, it's interesting
to see $\frac{j_{3}^{[N]}}{\frac{j_{2}^{[N]}}{\frac{j_{1}^{[N]}}{\Lambda_{n-1}}-\Lambda_{n-1}}+\Lambda_{n-1}}$
and $\frac{j_{2}^{[N-2]}}{\frac{j_{1}^{[N-2]}}{\Lambda_{n-2}}-\Lambda_{n-2}}$
have the same structure when a similar replacement as in even $N$
cases been made. With the help of this property, we can prove $\Lambda_{n-1}$
is also an eigenvalues of $H_{N}$. Combining it with the fact that
$\Lambda_{n}$ is an eigenvalue of $H_{N}$ and the symmetry property
between $\Lambda_{n}$ and $\Lambda_{n-2}$, by means of mathematical
induction, we assert the eigenvalues of $H_{N}$ whose off-diagonal
elements are constructed from Eq.(\ref {10}) are $\{0,\pm\Lambda_{i}\}$
for $i=1,\ldots,n$ indeed.

The completeness of this method comes from the fact that \eqref{HMatrix}
is uniquely determined by its eigenvalues when $\{J_{i}\}$ are all
positive\cite{Hochstadt74LAA}. This also implies all real coupling
schemes for \eqref{HN} are uniquely determined by its eigenvalues.
Since the completeness is available only if all $\{J_{i}\}$ are real,
we need to prove the positivity of $\{J_{i}^{2}\}$ for Eq.\eqref{EvenRec}
and Eq.(\ref {10}). This is more apparent when we factor out the
common factors of each equation in Eq.\eqref{EvenRec} and Eq.(\ref {10}).
After factorization, we will find each expression contains two factors,
one is positive and the other is monotone with respect to $\Lambda_{n}$.
Considering $\Lambda_{n-1}>\Lambda_{n}>0$ and $\Lambda_{n}>\Lambda_{n-1}$
in even and odd cases respectively, we assert $\{j_{n}^{[N]}\}$ are
all positive and $\{J_{i}\}$ are all real which satisfy the completeness
condition. In a word, Eq.\eqref{EvenRec} and Eq.(\ref {10}) are
complete for all possible coupling schemes.

Up to now, we have solved both even and odd $N$ cases in the absence
of external control fields $\{B_{i}\}$. This constructive method
allows us to calculate the couplings from a set of preselected eigenvalues.
We have chosen several sets of $50$ numbers whose interval between
any two adjacent ones in each set is a random odd number in the domain
$[1,100]$. In general, we got the couplings within
10 seconds. This numerical calculation shows our method is effective.
Although the resultant couplings often have enormous numerators and
denominators caused by the continued fraction structure of the constructive
method, we can choose some specific eigenvalues and then get compact
coupling schemes. For example, choosing the eigenvalues as $\{\pm(T+\frac{1}{2}+i(2S+1))\}$,
where $T$ and $S$ are two non-negative integers, for $i=1,2\ldots,\frac{N}{2}$
when $N$ is even, we will find $J_{i}^{2}$ are $\frac{i(N-i)(1+2S)^{2}}{4}$
and $\frac{((1+2T)+(1+i)(1+2S))((1+2T)+(N+1-i)(1+2S))}{4}$ for even and
odd $i$ respectively.

The model used here is also similar to that we encounter in population
transfer in an $N$-level system in which $N$ discrete energy levels
are equivalent to $N$ single excited states $\{\left|i\right\rangle \}$.
Assuming the only interaction to be that of electric-dipole transitions
and each frequency of the laser to be close to resonance with two
adjacent states, after rotating wave approximation (a general review
of this topic, see\cite{Shore08ActaPhysSlovaca}), Hamiltonian of
this problem is identical to Eq.\eqref{HN} when we treat the dipole
interactions as the couplings in $XY$ chain. Our results for PST
can also be used to design the amplitude of each frequency of the
control laser to achieve perfect population transfer. 

In this Letter, we have considered the problem of transferring an
unknown state from one end of a spin chain to the other end, and proposed
two recursive formulas for designing the couplings since uniform coupled
$XY$ chains can not afford PST. We also prove these formulas are
complete. Although this method is numerically effective, there are
still some interesting issues. We set the diagonal elements to be
zeros, i.e. there is no external control field in spin chain or the
laser resonances with any two adjacent levels in an $N$-level system.
This is not necessary for PST or perfect population transfer. Non-zero
diagonal elements break the symmetry of the spectrum of the Hamiltonian,
and the eigenvalues no longer appear in pairs. Nevertheless, the continued
fraction is also available when we replace $\Lambda$ by $B_{i}-\Lambda$.
We expect similar formula for cases involve control fields which,
of course, will contain $N$ recursive equations but not $[N/2]$
for an $N$-site spin chain. Another question is whether there are
other simple coupling schemes for special selected eigenvalues. We
have tested some simple sets of eigenvalues, but the couplings still
seem complicated.

This work was supported in part by Foundation of President of Hefei
Institutes of Physical Science CAS, one of us (F.S.) was also partly
supported by National Natural Science Foundation of China (No. 61074052).

\bibliographystyle{apsrev}

\end{document}